# A Portfolio-Level Optimization Framework for Coordinated Market Participation and Operational Scheduling of Hydrogen-Centric Companies

Seyed Amir Mansouri[1] and Kenneth Bruninx[1]

[1] Department of Engineering Systems & Services, Faculty of Technology, Policy & Management, Delft University of Technology, Delft, The Netherlands

s.mansouri@tudelft.nl and k.bruninx@tudelft.nl

*Abstract*— **The vision of electrolytic hydrogen as a clean energy vector prompts the emergence of hydrogen-centric companies that must simultaneously engage in electricity, hydrogen, and green certificate markets while operating complex, geographically distributed asset portfolios. This paper proposes a portfolio-level optimization framework tailored for the integrated operational scheduling and market participation of such companies. The model co-optimizes asset scheduling and market decisions across multiple sites, incorporating spatial distribution, technical constraints, and company-level policy requirements. It supports participation in the electricity market, physical and virtual Power Purchase Agreements (PPAs), bundled and unbundled hydrogen markets, and green certificate transactions. The model is applied to three operational scenarios to evaluate the economic and operational impacts of different compliance strategies. Results show that centralized, portfolio-level control unlocks the full flexibility of geographically distributed assets, enabling a 2.42-fold increase in hydrogen production and a 9.4% reduction in daily operational costs, while satisfying all company policy constraints.**

*Keywords*— *Green Hydrogen; Clean Electricity; Multi-Market Scheduling; Sector Coupling; Power Purchase Agreements.*

## I. INTRODUCTION

The global commitment to decarbonization will require accelerating the deployment of hydrogen as a flexible, clean energy vector. Hydrogen is increasingly viewed as a strategic enabler for deep electrification, sector coupling, and large-scale energy storage. In particular, green hydrogen produced via renewable-powered electrolysis is emerging as a critical pillar in achieving net-zero targets, especially in sectors that are hard to electrify directly [1]. In this evolving landscape, hydrogen-centric companies, i.e., organizations integrating electrolyzers, renewable energy assets, and storage systems, are positioned to play a vital role in the development of future energy systems. These companies must navigate complex operational environments, participate in multiple energy markets, and comply with evolving regulatory frameworks. Their success hinges on their ability to manage a geographically distributed asset portfolio, operate flexibly in response to market signals, and strategically coordinate their participation across electricity, hydrogen, and green certificate markets. The design of robust operational and market strategies for such actors is not only essential for their own economic viability, but also for ensuring that hydrogen integration contributes meaningfully to system-wide efficiency, flexibility, and sustainability. As electricity systems become more variable and decentralized, hydrogen-centric business models are uniquely positioned to offer grid services, absorb excess renewable generation, and provide cross-vector balancing capabilities. This paper is motivated by the pressing need to support emerging hydrogen-centric companies with advanced optimization frameworks capable of capturing the complexity of interconnected energy markets and multi-asset operations. We present an optimal scheduling problem considering hydrogen-centric companies, integrating asset-level constraints, company-level policy requirements, and the coupling of electricity, hydrogen, and green certificate markets.

### A. Related Works

Operational scheduling of hydrogen-based prosumers is a key enabler for integrating hydrogen technologies into decentralized multi-energy systems. In this context, [2] proposed a tri-objective scheduling model for renewable-hydrogen microgrids with storage and demand-side management, optimizing economic, environmental, and resilience objectives. [3] applied an Information Gap Decision Theory (IGDT)-based framework for day-ahead scheduling of energy communities with shared hydrogen infrastructure, ensuring robust prosumer participation under demand and price uncertainty. [4] analyzed economic-environmental trade-offs in industrial multi-energy systems, emphasizing price-responsive hydrogen scheduling. [5] extended this by incorporating biomass-hydrogen systems into a multi-objective hub model using IGDT to manage renewable and storage uncertainties. [6] introduced a two-stage reinforcement learning approach for hydrogen fueling station scheduling, aligning operations with grid constraints while enhancing safety. Lastly, [7] presented a risk-based scheduling method using water electrolysis and methane reforming, supporting cost-effective and integrated energy system operation.

The evolving energy market landscape requires hydrogen-based entities to engage across electricity, hydrogen, and environmental attribute markets to optimize uncertain economic returns. In this context, [8] investigates green certificate allocation within hydrogen supply chains, demonstrating its efficacy in reducing costs and enabling dispatch flexibility. National policy frameworks and financial support mechanisms also play a pivotal role in shaping the development of the green hydrogen market, as highlighted by [9]. The authors in [10] recommend coordinated, multi-market participation by hydrogen suppliers to address rising demand and market volatility. A detailed pricing model presented in

[11] for hydrogen refueling infrastructure underscores the importance of incorporating market signals and cost-reflective tariffs. From a distributed generation perspective, [12] demonstrate that green certificate mechanisms can incentivize Photovoltaic (PV) adoption, thereby increasing the availability of renewable electricity for hydrogen production. Furthermore, [13] propose a strategic bidding model for virtual power plants participating in electricity, carbon, and certificate markets, showing that regulatory coupling and dynamic pricing can simultaneously enhance revenue streams and contribute to emissions reduction. PPAs have emerged as a key mechanism for scaling up green hydrogen by ensuring stable, long-term access to clean electricity. [14] presents a techno-economic model indicating that PPAs improve the bankability of hydrogen projects by hedging against spot market volatility. [15] propose a modified PPA structure for urban decarbonization that aligns production with renewable availability and regulatory targets. Complementing these insights, [16] show that PPAs enable temporal alignment between generation and electrolyzer operation, essential for certification compliance.

### B. Research Gaps and Contributions

Despite the growing body of research on hydrogen integration within multi-energy systems, existing models remain predominantly constrained to asset-level operations and/or isolated market segments. While recent studies have advanced dispatch optimization for electrolyzer-based systems, yet typically treat electricity, hydrogen, and certificate markets as disjointed markets, with limited attention to their interdependencies. Moreover, critical factors such as the spatial distribution of assets and company-level policy constraints and the distinction between bundled and unbundled products, are often either overlooked or handled with excessive simplification. To address the identified gaps, this paper presents a centralized, portfolio-level optimization framework that integrates electricity, hydrogen, and certificate markets; incorporates spatial and regulatory complexities, including additionality requirements and product unbundling distinctions; and enables coordinated decision-making across distributed assets in hydrogen-centric companies. The key contributions are:

- Developing a portfolio-level optimization model that coordinates the scheduling of distributed energy assets across multiple sites, enabling coordination beyond individual operations and unlocking flexibility. This model allows co-optimizing participation across electricity, hydrogen (bundled and unbundled), and green certificate markets, while rigorously integrating company-level policy targets, such as a green hydrogen target.
- Analyzing the impact of company-level policy constraints and PPA dispatch formats on multi-market costs, the model shows that centralized, portfolio-level coordination reduces operational costs and boosts hydrogen production, while ensuring compliance across electricity, hydrogen, and certificate markets.

The remainder of the paper is organized as follows. Section II presents the model outline. The mathematical formulation is presented in Section III. Before concluding (Section V), a case study is discussed in Section IV.

## II. MODEL OUTLINE

The model's primary objective is to establish a comprehensive optimization structure for both the operational planning and market strategy formulation of hydrogen-centric companies. The companies under consideration integrate electrolyzers, renewable energy sources, and Electrical Energy Storage (EES) systems, which are geographically distributed across a wide region (see Section IV). At the core of the proposed optimization model lies a central control unit, representing the company's decision-making hub; an overview of this architecture is depicted in Fig. 1. This unit is responsible for coordinating the operation of all company-owned assets, dispatching energy under contracted PPAs, and determining the level of market participation across electricity, hydrogen, and green certificate markets. Through this centralized architecture, the model supports the coordination and alignment of market and operational decisions in a unified manner, accommodating company-level policy constraints, temporal interdependencies across electricity, hydrogen, and green certificate markets, and the technical characteristics of each asset. A key feature of the model is its ability to observe all assets simultaneously, thereby unlocking the maximum available flexibility across the asset portfolio to enhance the company's overall profitability. As depicted in the model overview, the control unit has access to both physical and virtual PPAs, which are assumed to follow a 'take-as-produced' contractual format. The company receives Guarantees of Origin (GOs) certificates corresponding to the output levels specified in these PPAs. Moreover, the company is modeled as an active participant in power, hydrogen, and GO markets, all operating within a day-ahead framework. In the context of hydrogen trading, the company engages in both bundled hydrogen transactions, hydrogen paired with GOs, and unbundled transactions representing hydrogen without certification. The proposed model enables integrated portfolio-level operational planning and coordinated participation in electricity, hydrogen, and certificate markets, while ensuring compliance with company-level policy targets.

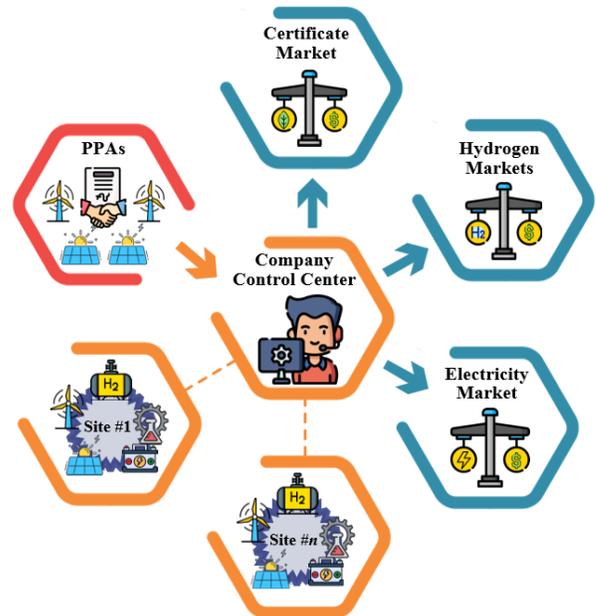

Fig. 1: Proposed Model for Hydrogen-Centric Companies

### III. MODEL FORMULATION

The proposed day-ahead operational planning framework is formulated as a Mixed-Integer Linear Programming (MILP) model for hydrogen-centric companies. The company operates a geographically distributed portfolio comprising wind and solar units, EES, and electrolyzer-based hydrogen production. Electricity for electrolysis is sourced from on-site renewables, the day-ahead market, or PPAs. Hydrogen is either consumed in Downstream Processes (DSPs) or sold in bundled or unbundled markets. The company also trades in the green certificate market to meet company targets or monetize surplus GOs. Both physical and virtual PPAs are modeled under a take-as-produced structure: physical PPAs involve direct electricity delivery with no financial settlement, while virtual PPAs serve as financial hedges, with payments based on the spread between the market and contractual prices. In both cases, GOs are transferred to the buyer to certify renewable origin.

The objective function of the company's planning model, expressed in Eq. (1), seeks to maximize total profit and includes five components: (i) profit from hydrogen sales in both bundled and unbundled markets; (ii) the net cost or revenue associated with GO transactions; (iii) the net cost or revenue of electricity exchanges in the day-ahead market; (iv) total costs under physical PPAs, including the cost of delivered electricity, grid access and transmission charges, and the cost associated with energy losses during delivery; and (v) the contract-for-difference settlements associated with virtual PPAs. Equation (2) specifies that if the market price ($\lambda_t^{DAM-Buy}$) falls below the contractual price ($\lambda^{VPPA}$), the off-taker compensates the generator for the difference; conversely, if the market price exceeds the contractual price, the generator compensates the off-taker for the surplus.

$$\max z^{H2FLEX} = \sum_t \begin{bmatrix} \left( \lambda^{BD} h_t^{BD,H2FLEX} + \lambda^{UBD} h_t^{UBD,H2FLEX} \right) \\ + \begin{pmatrix} \lambda^{GOs-Sell} n_t^{GOs-Sell,H2FLEX} \\ -\lambda^{GO-Buy} n_t^{GOs-Buy,H2FLEX} \end{pmatrix} \\ + \begin{pmatrix} \lambda_t^{DAM-Sell} p_t^{DAM-Sell,H2FLEX} \\ -\lambda_t^{DAM-Buy} p_t^{DAM-Buy,H2FLEX} \end{pmatrix} \Delta t \\ - \begin{pmatrix} (1+\alpha^{Loss})\lambda^{PPPA} P_t^{PPPA,H2FLEX} \\ +\lambda^{GA\&TR} P_t^{PPPA,H2FLEX} \end{pmatrix} \Delta t \\ -c_t^{CfD} \end{bmatrix} \quad (1)$$

$$c_t^{CfD} = \left( \lambda^{VPPA} - \lambda_t^{DAM-Buy} \right) P_t^{VPPA,H2FLEX} \Delta t \quad (2)$$

The hydrogen production at each electrolyzer is defined by Eq. (3), where $p_{e,t}^{EL}$ and $\eta_e^{EL}$ represent electricity consumption and efficiency of electrolyzer $e$. The asset-level hourly power balance is formulated in Eq. (4), power injected by local renewable units ($p_{e,t}^{Local}$), power delivered through physical PPAs ($p_{e,t}^{PPPA}$), net exchanges with the day-ahead market, net operations of the EES ($p_{e,t}^{EES}$), and power consumption by the electrolyzer. Equation (5) ensures that power injected from local units is supplied exclusively by wind and PV sources. The operational constraint in Eq. (6) enforces that the electrolyzer's operating point must be no lower than the level required to meet DSP hydrogen demand. Equation (7) determines the hydrogen consumed in the DSP at each site, as a function of the electrolyzer capacity $P_e^{EL,Max}$ and operational flexibility rate $\alpha_e^{Flex}$. Constraints (8)-(11) aggregate asset-level variables to the portfolio level: Eqs. (8) and (9) relate to physical and virtual PPA transactions, while Eqs. (10) and (11) capture the exchanges with the electricity market. Constraints (12)-(14) ensure the mutual exclusivity between power purchased from and sold to the market. $M$ denotes a sufficiently large constant, whereas $i_{e,t}^{DAM-Buy}$ and $i_{e,t}^{DAM-Sell}$ are binary variables.

$$h_{e,t}^{EL} = \eta_e^{EL} p_{e,t}^{EL} \quad (3)$$

$$p_{e,t}^{Local} + p_{e,t}^{PPPA} + (p_{e,t}^{DAM-Buy} - p_{e,t}^{DAM-Sell}) = p_{e,t}^{EES} + p_{e,t}^{EL} \quad (4)$$

$$p_{e,t}^{Local} = F_t^{Wind} P_e^{Wind,Max} + F_t^{PV} P_e^{PV,Max} \quad (5)$$

$$\frac{h_{e,t}^{DSP}}{\eta_e^{EL}} \leq p_{e,t}^{EL} \leq P_e^{EL,Max} \quad (6)$$

$$h_{e,t}^{DSP} = \eta_e^{EL} P_e^{EL,Max} (1-\alpha_e^{Flex}) \quad (7)$$

$$\sum_e p_{e,t}^{PPPA} = P_t^{PPPA,H2FLEX} \quad (8)$$

$$\sum_e p_{e,t}^{VPPA} = P_t^{VPPA,H2FLEX} \quad (9)$$

$$p_t^{DAM-Buy,H2FLEX} = \sum_e p_{e,t}^{DAM-Buy} \quad (10)$$

$$p_t^{DAM-Sell,H2FLEX} = \sum_e p_{e,t}^{DAM-Sell} \quad (11)$$

$$p_{e,t}^{DAM-Buy} \leq i_{e,t}^{DAM-Buy} M \quad (12)$$

$$p_{e,t}^{DAM-Sell} \leq i_{e,t}^{DAM-Sell} M \quad (13)$$

$$i_{e,t}^{DAM,Buy} + i_{e,t}^{DAM,Sell} \leq 1 \quad (14)$$

It is assumed that each site is equipped with an EES system co-located with its renewable units. The state-of-charge of the EES is defined in Eq. (15), which considers the previous state-of-charge along with the charging and discharging actions during the current time step. The maximum allowable charging and discharging rates per hour are constrained by Eqs. (16) and (17). Constraint (18) prohibits simultaneous charging and discharging. The net hourly power exchanged with the EES, either injected into or withdrawn from the system, is represented by variable $p_{e,t}^{EES}$ in Eqs. (19). Constraint (20) ensures that the energy level remains within predefined lower and upper bounds. Note that the energy levels of the EES at the initial and final hours are set to 50% of its rated capacity.

$$e_{e,t}^{EES} = e_{e,t-1}^{EES} + \eta_e^{EES-Ch} p_{e,t}^{Ch} \Delta t - \frac{p_{e,t}^{Dis}}{\eta_e^{ESS-Dis}} \Delta t \quad (15)$$

$$0 \leq p_{e,t}^{Ch} \leq P_e^{Ch,Max} i_{e,t}^{EES-Ch} \quad (16)$$

$$0 \leq p_{e,t}^{Dis} \leq P_e^{Dis,Max} i_{e,t}^{EES-Dis} \quad (17)$$

$$i_{e,t}^{EES-Ch} + i_{e,t}^{EES-Dis} \leq 1 \quad (18)$$

$$p_{e,t}^{EES} = p_{e,t}^{Ch} - p_{e,t}^{Dis} \quad (19)$$

$$E_e^{EES,Min} \leq e_{e,t}^{EES} \leq E_e^{EES,Max} \quad (20)$$

The hourly asset-level hydrogen balance is defined by Eq. (21), incorporating hydrogen produced by the electrolyzer, hydrogen sold in the bundled and unbundled markets, and hydrogen consumed in the DSP. Equations (22) and (23) aggregate the asset-level hydrogen sale variables to the portfolio level.

$$h_{e,t}^{EL} = h_{e,t}^{BD} + h_{e,t}^{UBD} + h_{e,t}^{DSP} \quad (21)$$

$$h_{e,t}^{BD,H2FLEX} = \sum_e h_{e,t}^{BD} \quad (22)$$

$$h_{e,t}^{UBD,H2FLEX} = \sum_e h_{e,t}^{UBD} \quad (23)$$

Equation (24) requires that the total number of certificates acquired, via market purchases, PPAs, and local renewable units, matches the total number sold to the market, allocated to the DSP, and assigned to bundled hydrogen sales. Equations (25)-(27) compute the certificates received from virtual PPAs, physical PPAs, and local renewable units,. The conversion factor $\gamma^{GOs}$, set at 1 unit/MW, represents the electricity-to-GOs certificate ratio. Equations (28) and (29) compute the certificates assigned to bundled hydrogen sales and to the DSP.

$$\sum_t \left( n_{e,t}^{GOs-VPPA} + n_{e,t}^{GOs-PPPA} + n_{e,t}^{GOg-Local} + n_{e,t}^{GOs-Buy} \right)$$
$$= \sum_t \left( n_{e,t}^{GOs-Sell} + n_{e,t}^{GOs-BD} + n_{e,t}^{GOs-DSP} \right) \quad (24)$$

$$n_{e,t}^{VPPA} = \gamma^{GOs} P_{e,t}^{VPPA} \quad (25)$$

$$n_{e,t}^{PPPA} = \gamma^{GOs} P_{e,t}^{PPPA} \quad (26)$$

$$n_{e,t}^{GOs-Local} = \gamma^{GOs} p_{e,t}^{Local} \quad (27)$$

$$n_{e,t}^{GOs-BD} = \gamma^{GOs} \frac{h_{e,t}^{BD}}{\eta_e^{EL}} \quad (28)$$

$$n_{e,t}^{GOs-DSP} = \gamma^{GOs} \frac{h_{e,t}^{DSP}}{\eta_e^{EL}} \quad (29)$$

The company-level green hydrogen target is defined in Eq. (30), enforcing that at least 90% of the total daily hydrogen production is covered by GOs, using a compliance factor $\alpha^{Green}$. Constraints (31)-(33) prevent the simultaneous purchase and sale of GOs in the green certificate market. Equations (34) and (35) map asset-level variables related to GOs exchanges to portfolio-level variables. Finally, Eqs. (36) and (37) constrain the total daily volume of GOs that can be purchased from and sold to the market.

$$\sum_e \left[ \frac{1}{\gamma^{GOs}} \sum_t \left( \begin{array}{c} n_{e,t}^{GOs-VPPA} + n_{e,t}^{GOs-PPPA} \\ + n_{e,t}^{GOs-Local} + n_{e,t}^{GOs-Buy} - n_{e,t}^{GOs-Sell} \end{array} \right) \right]$$
$$\geq \alpha^{Green} \times \sum_e \left[ \frac{1}{\eta_e^{EL}} \sum_t \left( h_{e,t}^{BD} + h_{e,t}^{UBD} + h_{e,t}^{DSP} \right) \right] \quad (30)$$

$$n_{e,t}^{GOs-Sell} \leq i_{e,t}^{GOs-Sell} M \quad (31)$$

$$n_{e,t}^{GOs-Buy} \leq i_{e,t}^{GOs-Buy} M \quad (32)$$

$$i_{e,t}^{GOs-Sell} + i_{e,t}^{GOs-Buy} \leq 1 \quad (33)$$

$$n_t^{GOs-Sell,H2FLEX} = \sum_e n_{e,t}^{GOs-Sell} \quad (34)$$

$$n_t^{GOs-Buy,H2FLEX} = \sum_e n_{e,t}^{GOs-Buy} \quad (35)$$

$$\sum_e \sum_t n_{e,t}^{GOs-Buy} \leq N^{GOs-Buy,Max} \quad (36)$$

$$\sum_e \sum_t n_{e,t}^{GO-Sell} \leq N^{GOs-Sell,Max} \quad (37)$$

## IV. SIMULATION RESULTS

The proposed optimization model is implemented in Python using Pyomo, with input data and code publicly available in [17]. The wind and solar generation profiles, along with market prices, are based on data from a representative day in March 2025 in Spain. As illustrated in Fig. 2, the model is applied to H2FLEX, a hydrogen-centric company operating five sites distributed across a broad geographical area, evaluated under three distinct operational setups: **(i)** Case 1, where each electrolyzer operates independently with its own PPA and individual, uniform green hydrogen target constraints; **(ii)** Case 2, where PPAs are centrally dispatched among electrolyzers by the main company, while company-level green hydrogen targets are still enforced on each site individually; and **(iii)** Case 3, where both PPAs and green hydrogen targets are managed at the portfolio level by the company. All cases are modeled over a representative day with hourly resolution. In Case 1, constraint (8) and the green hydrogen targets (30) are enforced per site. Case 2 applies constraint (8) at the portfolio level, with other settings unchanged. Case 3 shifts constraint (30) to the portfolio level, with all else as in Case 2.

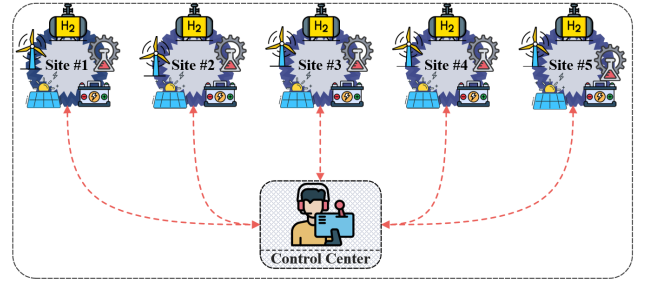

Fig. 2: Overview of Hydrogen-Centric Company Under Study

Table I presents the numerical results on a daily basis for the three operational scenarios evaluated. The numerical results show identical PPA transactions across all cases due to their take-as-produced structure, which obliges the company to accept and schedule the full volume of electricity generated under the PPA, regardless of local demand or operational flexibility. As a result, the contracted energy must either be consumed on-site, stored in EES systems, or sold in the market. Meanwhile, market participation varies, with Case 2 reducing daily net costs by 7.9% and Case 3 by 9.4%, both relative to Case 1. The cost reduction in Case 2 is primarily attributed to the centralized management of the PPAs, which enables the company to allocate PPA-sourced electricity across geographically distributed sites in a coordinated manner. In contrast, Case 1 assumes that each site independently manages its own PPA without sharing resources, thereby limiting the overall system flexibility and leading to suboptimal asset utilization. Notably, in Case 1, a total of 40.5 units of GOs were sold in the certificate market, whereas in Case 2, no GOs were sold. Instead, all certificates were strategically retained to support bundled hydrogen sales. This internal reallocation led to a 1161.3% (811.8 kg) increase in bundled hydrogen sales. Furthermore, unbundled hydrogen sales also rose by 6.8% (90.2 kg), from 1314.4 kg in Case 1 to 1404.6 kg in Case 2, highlighting the improved operational flexibility facilitated by centralized coordination.

TABLE I. SUMMARY OF SIMULATION OUTCOMES FOR CASES 1-3

| Case | Virtual PPAs | | Physical PPAs | | Electricity Market | | | Bundled H$_2$ Market | | Unbundled H$_2$ Market | | Certificate Market | | | Green H$_2$ (%) | Sum (€) |
|---|---|---|---|---|---|---|---|---|---|---|---|---|---|---|---|---|
| | Qty (MWh) | CfD (€) | Qty (MWh) | Costs (€) | Buy (MWh) | Sell (MWh) | Net (€) | Qty (kg) | Rev. (€) | Qty (kg) | Rev. (€) | Buy (unit) | Sell (unit) | Net (€) | | |
| 1 | 249.7 | 6571.9 | 642.3 | 38862.8 | 536.2 | 41.3 | 13501.8 | 69.9 | 409.2 | 1314.4 | 2628.8 | 220 | 40.5 | 937.6 | 94.2 | 56836.1 |
| 2 | 249.7 | 6571.9 | 642.3 | 38862.8 | 539.9 | 0 | 13732.4 | 881.7 | 5158.2 | 1404.6 | 2809.2 | 220 | 0 | 1100 | 94 | 52299.7 |
| 3 | 249.7 | 6571.9 | 642.3 | 38862.8 | 593.2 | 0 | 15044.4 | 881.7 | 5158.2 | 2471.3 | 4942.6 | 220 | 0 | 1100 | 90 | 51478.3 |

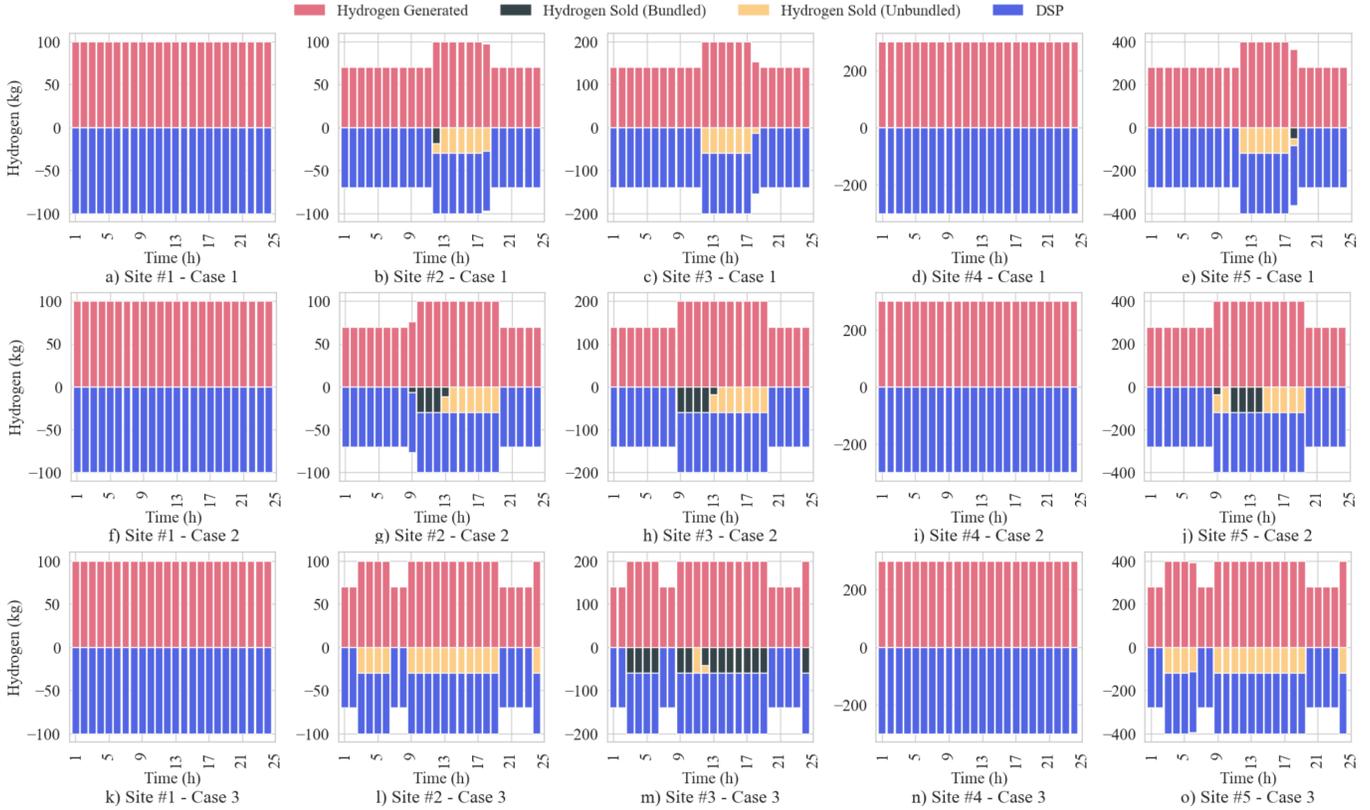

Fig. 3: Operational Schedules Across Sites for Evaluated Scenarios

The schedules illustrated in Fig. 3 further validate these findings. In both Cases 1 and 2, the inflexible sites (Sites 1 and 4) maintain the same operation, using all PPA-sourced electricity to supply DSPs. In contrast, the flexible sites (Sites 2, 3, and 5) adjust their schedules in Case 2 to boost hydrogen production and enhance participation in both bundled and unbundled markets. In Case 1, Site 1's isolated PPA management results in 41.3 MWh of excess electricity being sold to the market (not shown in Fig. 3). However, in Case 2, coordinated PPA dispatch reallocates surplus to other sites for hydrogen production, avoiding exports and boosting revenues via greater market participation. These results confirm that centralized portfolio-level dispatch of PPAs enables more flexible and cost-effective use of clean electricity across sites. It unlocks system-wide optimization potential, enhancing hydrogen production, compliance with certification requirements, and economic performance.

Further improvements are observed in Case 3, where the company-level green hydrogen target is enforced at the portfolio level rather than on a per-site basis. This configuration results in a 1.5% reduction in total company costs compared to Case 2. By leveraging aggregated operational flexibility, the company is able to increase hydrogen production by 46.6%, leading to greater participation in the unbundled hydrogen market. Although the percentage of green hydrogen in the company's portfolio decreases from 94% in Case 2 to exactly 90% in Case 3, this reduction reflects an intentional strategy to maximize the permitted share of gray hydrogen production, thereby optimizing economic performance while remaining within company-limits. Under portfolio-level compliance, the company can precisely allocate 10% of total hydrogen production to gray sources, whereas asset-level enforcement in

Case 2 constrains this flexibility, leading to under-utilization of available production capacity. The dispatch profiles in Fig. 3 for Case 3 confirm that the scheduling of flexible sites (Sites 2, 3, and 5) differs markedly from Case 2. In Case 3, all hydrogen sold at Sites 2 and 5 is allocated exclusively to the unbundled market, whereas in Case 2, sales at these sites were split between bundled and unbundled markets. Moreover, all hydrogen sales to the bundled market are concentrated at Site 3. The results show that centralized PPA dispatch and portfolio-level enforcement of the additionality constraint, rather than at the asset level, unlock the full operational flexibility of distributed assets, yielding a more optimal internal schedule and increasing overall profitability via improved market participation.

## V. Conclusion

This paper proposed a portfolio-level optimization framework to support the operational planning and market participation of hydrogen-centric companies across electricity, hydrogen, and green certificate markets. The model integrated asset-level constraints, company-level policy requirements, and market interactions, enabling coordinated scheduling of distributed assets. It captured the contractual heterogeneity of physical and virtual PPAs, differentiated between bundled and unbundled hydrogen products, and enforced additionality constraints for green hydrogen certification. The simulation results demonstrated that centralized PPA dispatch and portfolio-level coordination of enforcing policy constraints significantly enhanced operational flexibility and economic performance. In our case study, centralized dispatch enabled a 7.9% reduction in operational costs and increased hydrogen sales through more effective internal allocation of electricity and green certificates. Moreover, enforcing additionality at the portfolio level, rather than at each site individually, further reduced costs by 1.5% and increased hydrogen production by 46.6%, all while maintaining full company policy compliance. Overall, the simulation outcomes testified that the proposed model served as a robust decision-support tool for hydrogen-centric companies operating in complex, multi-market environments. It enhanced the utilization of distributed asset flexibility, improved coordination with market mechanisms, and delivered quantifiable cost reductions.

## VI. Acknowledgement

This publication is part of the WinHy project (file number KICH2.V4P.DUI21.006) within the ECCM research program, which is (partly) funded by the Dutch Research Council (NWO) under grant no. 16663 and by Repsol S.A.